\documentclass[prl,superscriptaddress,aps, twocolumn, reprint, showpacs, longbibliography, footnoography]{revtex4-2}

\usepackage{graphicx}
\usepackage{amssymb}   
\usepackage{amsfonts}
\usepackage{amsmath}
\usepackage{dsfont}
\usepackage{natbib}
\usepackage{dcolumn}   
\usepackage{bm}        
\usepackage[mathscr]{eucal}
\usepackage[dvipsnames]{xcolor}
\usepackage[colorlinks, linkcolor=OrangeRed,citecolor=RoyalBlue,urlcolor=NavyBlue]{hyperref}
\usepackage[all]{hypcap} 
\usepackage{mathtools}
\usepackage{dsfont}

\definecolor{myblue}{rgb}{0,0,0.75}

\newcommand{\ep}{\varepsilon}

\newcommand\mean[1]{\ensuremath{\left\langle#1\right\rangle}}

\newcommand\lrp[1]{\left(#1\right)}
\newcommand\lrb[1]{\left[#1\right]}

\newcommand\lrv[1]{\left|#1\right|}

\newcommand{\be}{\begin{equation}}
\newcommand{\ee}{\end{equation}}
\def\ba{\begin{aligned}}
\def\ea{\end{aligned}}
\newcommand{\bea}{\begin{eqnarray}}
\newcommand{\eea}{\end{eqnarray}}
\def\bes{\begin{subequations}}
\def\ees{\end{subequations}}
\def\bal{\begin{align}}
\def\eal{\end{align}}

\renewcommand{\Re}{{\rm \, Re\,}}
\renewcommand{\Im}{{\rm \, Im\,}}
\renewcommand{\vec}[1]{{\bf #1}}

\newcommand{\rev}[1]{{#1}}
\newcommand{\revT}[1]{{#1}}

\usepackage{soul}
\usepackage{ulem} 

\usepackage{graphicx}
\usepackage{amsmath,amssymb,bm}
\usepackage{enumerate}
\usepackage{braket}        

\graphicspath{{./figures/}}
\begin{document}
\title{Non-Hermiticity induces localization: good and bad resonances in power-law random banded matrices}

\author{Giuseppe De Tomasi}
\affiliation{Department of Physics, University of Illinois at Urbana-Champaign, Urbana, Illinois 61801-3080, USA}

\author{Ivan M. Khaymovich}
\affiliation{Nordita, Stockholm University and KTH Royal Institute of Technology Hannes Alfv\'ens v\"ag 12, SE-106 91 Stockholm, Sweden}
\affiliation{Institute for Physics of Microstructures, Russian Academy of Sciences, 603950 Nizhny Novgorod, GSP-105, Russia}

\begin{abstract}
The power-law random banded matrix (PLRBM) is a paradigmatic ensemble to study the Anderson localization transition (AT). In $d$ \revT{dimensions}, the PLRBM are random matrices with algebraic decaying off-diagonal elements $H_{\vec{n}\vec{m}}\sim 1/|\vec{n}-\vec{m}|^\alpha$, having AT at $\alpha=d$. In this work, we investigate the fate of the PLRBM to non-Hermiticity (nH). We consider the case where the random on-site diagonal potential takes complex values, mimicking an open system, subject to random gain-loss terms. We understand the model analytically by generalizing the Anderson-Levitov resonance counting technique to the nH case. We identify two competing mechanisms due to nH: favoring localization and delocalization. The competition between the two gives rise to AT at $d/2\le \alpha\le d$. The value of the critical $\alpha$ depends on the strength of the on-site potential, like in Hermitian disordered short-range models in $d>2$. Within the localized phase, the wave functions are algebraically localized with an exponent $\alpha$ even for $\alpha<d$. This result provides an example of non-Hermiticity-induced localization \revT{and find immediate application in phase transition driven by weak measurements.}

\end{abstract}

\maketitle
\textit{Introduction}~--~Random matrix theory (RMT) is a resounding resource
to tackle problems in several contexts, ranging from nuclear physics to number theory~\cite{Terence_Tao_2011,mehta2004random, Beenakker_97, haake2001quantum}.
In disordered quantum systems, RMT has been widely used in Anderson localization (AL) and quantum chaos~\cite{haake2001quantum, Anderson_58, Evers2008Anderson, Beenakker_97, MIRLIN_2000}.
For instance, the Anderson transition (AT) has been successfully studied using the so-called power-law random banded matrix (PLRBM) ensemble~\cite{MirFyod1996} \revT{and its cousin ultrametric matrices (UM)~\cite{Fyodorov2009_UM} as remarkable examples}.
In $d$ dimensions, the PLRBM are \revT{matrices with random elements decaying} algebraically with the distance, $H_{\vec{n}\vec{m}}\sim \frac{1}{|\vec{n}-\vec{m}|^\alpha}$. 
\rev{The consideration }of the PLRBM 
in terms of resonance counting, resolved over spatial distance~\cite{Levitov1989}, renormalization group (RG)~\cite{Levitov1990}, and non-linear sigma model~\cite{MirFyod1996}, finding AT, at $\alpha_{AT}=d$, separating ergodic and localized phases, \rev{has boosted the field of AT far beyond this paradigmatic example.} That makes the PLRBM a fantastic toolbox for understanding universal properties in disordered quantum systems.
\revT{In addition, recently
the cousin UM model, having the same phase diagram and the properties, has found a direct application to the explanation of the avalanche mechanism of many-body localization (MBL) instability~\cite{Suntajs2023similarity}.
}
\begin{figure}[t!]
\label{fig:front_PLBM}
    \includegraphics[width=1.\columnwidth]{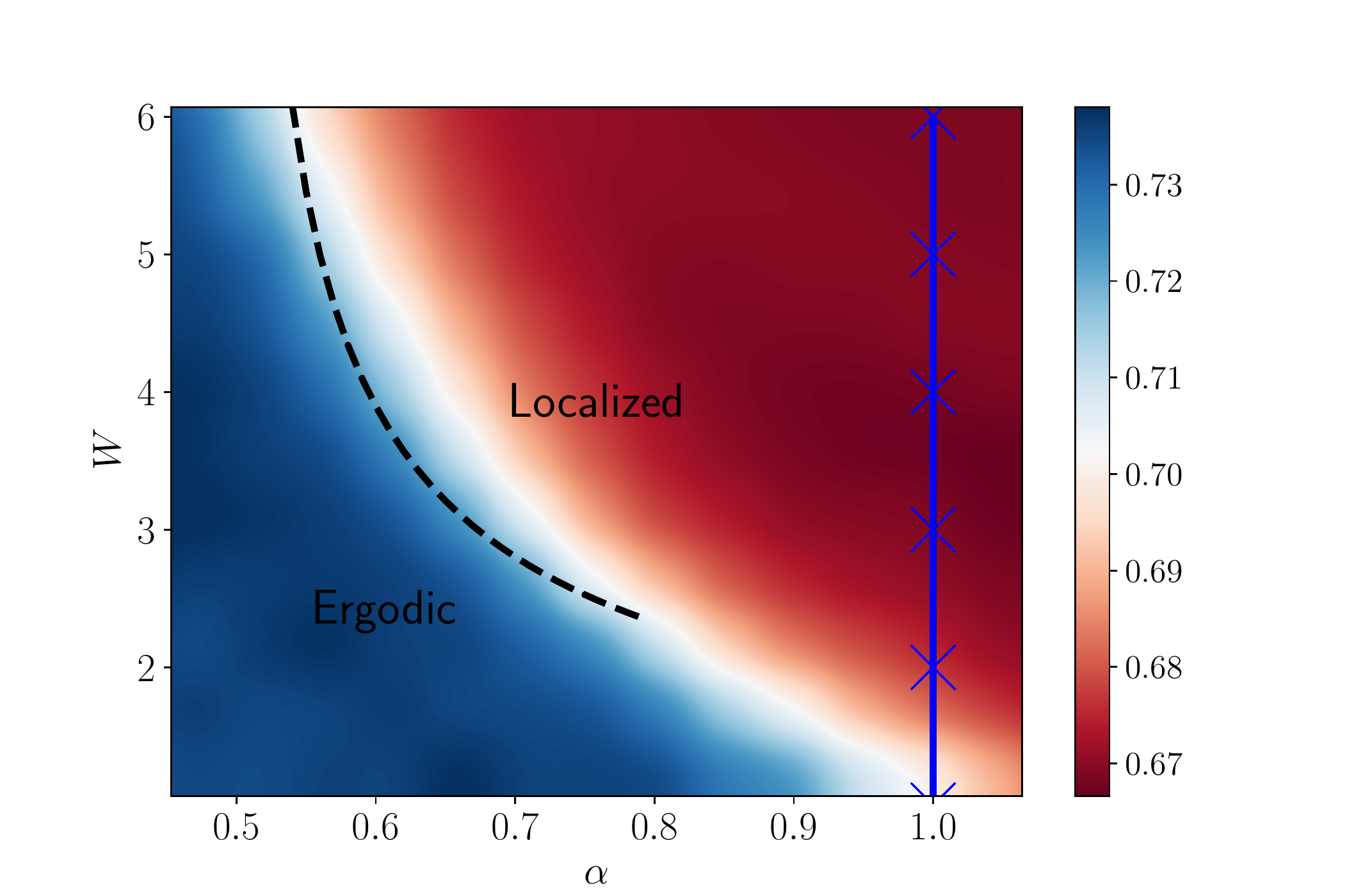}
    \caption{\textbf{Phase diagram of the non-Hermitian PLRBM with random gain-loss terms in $d=1$ dimension,} showing the averaged spectrum gap ratio $r$ vs the algebraic-decay rate $\alpha$ of the hopping terms and the diagonal disorder amplitude $W$. $\overline{r}=2/3$ for localized systems and $\overline{r}\approx 0.738$ for ergodic ones.
    For the Hermitian case, the transition happens at $\alpha=1$ for any $W$ (vertical blue line). The dashed line is the analytical prediction for AT.
    }
\end{figure}

\begin{figure*}[t!]
    \includegraphics[width=0.7\textwidth]{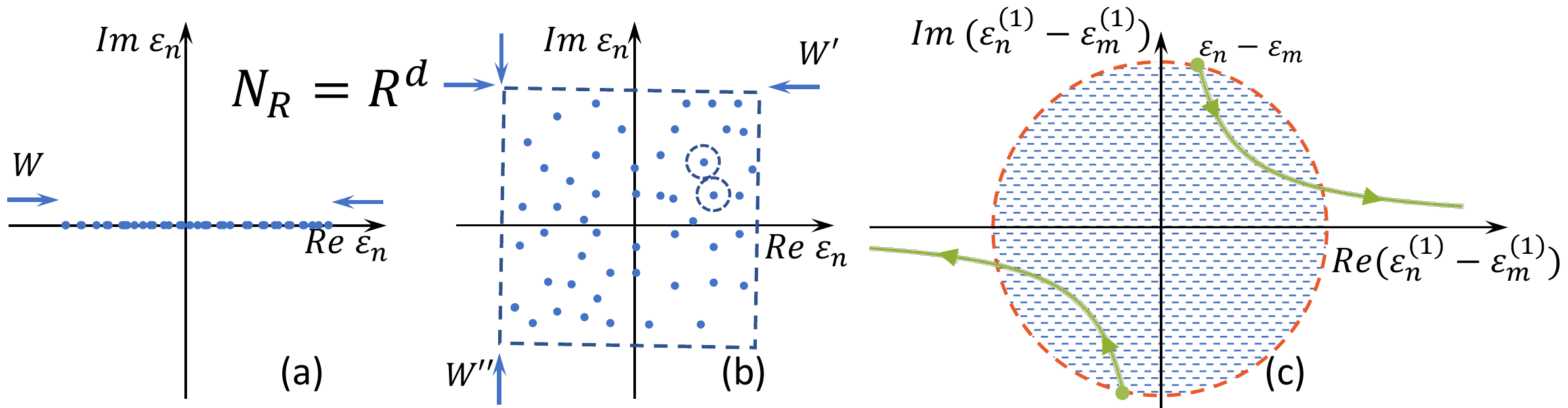}
    \caption{\textbf{Structure of resonances.}
    (a)~The level spacing in the Hermitian case, Eq.~\eqref{eq:delta_R}, is determined by the ratio of the energy interval $W$ and the number of lattice sites $N_R \sim R^d$ at the distance $R$ from a certain point. (b)~In the non-Hermitian case, the complex ($2$d) distribution of $\ep_n$ leads to the mean area per level $A_R\simeq W^2/R^d$, Eq.~\eqref{eq:delta_R_nH} and (c)~the hybridization of levels, Eq.~\eqref{eq:Em,En_expression}, may lead to the attraction between levels (shaded region), while the hopping term increases (green arrows).
    }
    \label{fig:nonHerm_resonances}
\end{figure*}

From a different perspective, non-Hermitian (nH) quantum systems have recently attracted significant attention. NH terms originated as an effective theory to describe system couples to baths and reservoirs. NH leads to a unique and broad phenomenology, ranging from skin effect~\rev{\footnote{\rev{Note that the skin effect, denoted as nH, has no relation to Anderson localization observed in transport phenomena. In Anderson localization, the left and right eigenvectors localize at opposite ends of the sample. At the same time, the transport properties are dictated by the overlap of these vectors' envelopes, which may still exhibit delocalization.}}}, measurement-induced transition, and generalized topology~\cite{Slager2020boundary,Okuma20,Yao_2018,Rudner_2009,Hu_2011,Esaki_2011,Gong_2018,Schomerus_13,Ashida2020,moiseyev_2011,Skinner_2019,Zabalo_2022,
Feinberg_1999,Molinari_2009,Huang_2020_ALT, Huang_2020_spectral,Ryu_2021,2021_Bergholtz_RevModPhys, Skinner_2019, Li_18, Chan_19}. In disordered systems, nH has drastic effects. In general, nH enhances decoherence and might break localization. The most well-known example is a so-called Hatano-Nelson model~\cite{Hatano_96} which shows that even in one dimension, where short-range Hermitian systems are localized for any amount of disorder, strong nH breaks AL. Consequently, nH  might fundamentally change the AT universality classes~\cite{Bernard2002_nonHerm_RMT_classes, Luo2021_Univers_classes,Luo2022_Unifying_nonHerm}.

\revT{Motivated by the above studies, i}n this work,
we take the route of studying AL in nH systems \revT{via their spatial resonance structure.}
In particular, we investigate a nH deformation of the PLRBM model. We break the Hermiticity by adding a complex on-site disorder but leaving Hermitian off-diagonal elements. Several works have considered this type of nH~\cite{Ueda_2019,Economou_2020, Huang_2020_ALT, Huang_2020_spectral, Luo2021_Univers_classes, Ryu_2021, Economou_2021, DeTomasi2022nonHerm_RP, DeTomasi2022nonHerm_MBL}, and this choice mimics local random gain-loss terms.

We show that Hermiticity-breaking terms change the phase diagram of the PLRBM, see Fig.~\ref{fig:front_PLBM}.
Using analytical and numerical techniques, we provide evidence that AT shifts to smaller values $d/2<\alpha_{c}(W)\le d$ and depends on the strength of the on-site disorder $W$.
To understand this counter-intuitive phenomenon, we generalize the Anderson-Levitov resonance counting and hybridization of resonant-level pairs, like in the standard RG~\cite{Levitov1989,Levitov1990}. We uncover the competition of two mechanisms due to nH, favoring localization and delocalization.
The complex-valued diagonal potential, unlike the real-valued one, Fig.~\ref{fig:nonHerm_resonances}(a), forms an effectively two-dimensional distribution~\cite{DeTomasi2022nonHerm_RP}, Fig.~\ref{fig:nonHerm_resonances}(b), which increases the level spacing parametrically in system size and suppresses the number of resonances. However, nH also induces a unique attraction between resonant levels, hybridizing them, Fig.~\ref{fig:nonHerm_resonances}(c), which forms the so-called ``bad'' resonances. The level attraction, caused by the nH terms, breaks the central assumption of spatial RG and dramatically changes the phase diagram. As a result, nH drives a new kind of transition that depends on the strength of on-site potential, like in short-range models in $d>2$. In addition, for the nH PLRBM, we show that the localized phase has an algebraic nature also for $\alpha \le d$, which is forbidden in its Hermitian counterpart.

\rev{Finally, we show that our results find immediate application in dynamical phase transitions, driven by the  competition between unitary time evolution and projective measurements, the so-called measurement-induced phase transitions. Indeed, the non-Hermitian Anderson transition is carried over to
a Floquet version of our system, in which the non-unitary part 
play a role of a weak measurement.
}

\textit{Model $\&$ Methods}~--~nH PLRBM in $d$ dimensions reads as
\be\label{eq:PLRBM_ham}
H_{\vec{mn}} = \ep_{\vec{n}} \delta_{\vec{mn}} + j_{\vec{mn}} \ ,
\ee
where $\ep_{\vec{n}}$ and $j_{\vec{mn}}=j^\star_{\vec{nm}}$ are independent complex-valued box-distributed random variables with zero mean and the amplitudes
\bes\label{eq:PLRBM_sigmas}
\begin{align}\label{eq:PLRBM_ep_n_complex}
|\Re\ep_{\vec{n}}|,|\Im\ep_{\vec{n}}| &\leq W, \\
\label{eq:PLRBM_j_mn_box}
|\Re j_{\vec{mn}}|^2, |\Im j_{\vec{mn}}|^2 &\leq \frac{1}{2(|\vec{m}-\vec{n}|^2+b^2)^\alpha} \ ,
\end{align}
\ees
$\vec{m}$ is a radial vector of a lattice in $\mathbb{Z}^d$, $W$ is the on-site disorder strength, $\alpha$~--~the power of the off-diagonal power-law decay, and $b$ is the bandwidth of the decay. For simplicity, we restrict our consideration to $d\leq 2$, where short-range models ($\alpha \to\infty$) are localized for any amount of disorder and take $b\lesssim 1$. The \revT{important gain-loss} nH is provided by the complex-valued diagonal term $\epsilon_{\vec{n}} \in \mathbb{C}$.
\rev{Furthermore, we consider an non-unitary Floquet version of Eq.~\eqref{eq:PLRBM_ham}, which in second-quantization is given by $U = e^{-i\revT{T}\sum_{\vec{mn}} j_{\vec{nm}} c^\dagger_\vec{m} c_\vec{n}} e^{-\revT{T}\sum_{\vec{n}} \ep_{\vec{n}} c^\dagger_\vec{n} c_\vec{n}}$\revT{, with a period $T$, set to be $1$ and creation (annihilation) fermionic operators $c^\dagger_\vec{n}$'s ($c_\vec{n}$'s).} The non-unitary part is given by $\Re\ep_{\vec{n}}$, which corresponds to weak-measurements~\cite{Skinner_2019,Chan_19,Li_18, Granet_22}. Two considerations are in order, the Floquet model is non-interacting and therefore preserves Gaussianity, allowing us to inspect its single-particle modes to understand its phase diagram, and the Anderson transition for Floquet Hermitian model happens at $\alpha_{AT} =1$~\cite{Burau_21}.}

The Hermitian PLRBM\revT{, as well as UM model~\cite{Fyodorov2009_UM},} hosts AT at $\alpha = d$, independent of the parameters $W$ and $b$. For $\alpha<d$, the wave function are extended and ergodic,
at the transition, the eigenstates show multifractality, weak for $b\gg 1$ and strong for $b \ll 1$~\cite{Evers2008Anderson}.
In the localized phase ($\alpha>d$), the wave functions are power-law localized, and 
the wave function decay rate coincides with $\alpha$. The transition is understood using the so-called Anderson-Levitov resonance counting~\cite{Levitov1989} with an RG approach~\cite{Levitov1990}. Its starting point is the perturbative \textit{locator expansion}
\be
\label{eq:locator}
\psi_{\vec{n}}(\vec{m}) = \delta_{\vec{mn}} + \frac{j_{\vec{mn}}}{\ep_{\vec{n}}-\ep_{\vec{m}}}+\ldots
\ee
which converges, if most of the site pairs $\vec{m}$ and $\vec{n}$ are not in resonance, $\left | \frac{j_{\vec{mn}}}{\ep_{\vec{n}}-\ep_{\vec{m}}} \right |<1$.
%
To find AT, one should count the number $N_{res,\vec{n}}$ of resonant pairs $\lrv{\frac{j_{\vec{mn}}}{\ep_{\vec{n}}-\ep_{\vec{m}}}}>1$ for each $\psi_{\vec{n}}$ state. For $N_{res,\vec{n}} \lesssim O(N^0)$ the perturbation theory is stable, and the state is localized, otherwise, for $N_{res,\vec{n}} \gtrsim O(N^{c>0})$ the state delocalizes.
At each $i$th step of the spatial RG~\cite{Levitov1989,Levitov1990}, when $|\vec{m}-\vec{n}|\equiv R$ is in a ring
\be\label{eq:PLRBM_R-ring}
R_i<R<2R_i\equiv R_{i+1} \ ,
\ee
one can count the number of resonances by comparing the mean-level spacing in the $d$-dimensional ring~\footnote{$\delta_R$ is given by the ratio
of the width of the on-site random potential $-W/2<\ep_{\vec{m}}<W/2$ to the number $\sim R^d$ of lattice sites in the ring~\eqref{eq:PLRBM_R-ring}}
\be\label{eq:delta_R}
\delta_R = \frac{W}{R^d} \ ,
\ee
with the hopping term $j_{\vec{m}\vec{n}}\sim 1/R^{\alpha}$
and progressively increase the distance at each step.
Thus, for all $\alpha>d$ and large enough distances $R\gtrsim(1/W)^{1/(\alpha-d)}\sim O(1)$ there are no resonances as $\delta_R/j_R = W R^{\alpha-d}> O(1)$.
$(1/W)^{1/(\alpha-d)}$ is the radius beyond which the resonances are absent. We can estimate the
number of sites resonant to $\vec{n}$ as $\sim (1/W)^{d/(\alpha-d)}$.
For these resonant sites, one must use degenerate perturbation theory. However, 
the hybridization of these resonance pairs does not produce other resonance pairs~\cite{Levitov1990}. As a result, for $\alpha>d$ the number of resonant pairs is bounded with system size.
Instead, $\alpha<d$ the resonances proliferate at large enough distances
$R>R_c\simeq W^{1/(d-\alpha)}$, as $j_R/\delta_R \sim R^{(d-\alpha)}/W > 1$.
The number of such resonances at a distance $R$ grows as
\be
N_{res,\vec{n}}(R) \simeq \frac{j_R}{\delta_R} \sim \frac{R^{d-\alpha}}{W} \ ,
\ee
leading to
\be
N_{res,\vec{n}} = \sum_i N_{res,\vec{n}}(R_i) \sim N^{d-\alpha} \gg N^0 \ .
\ee
This argument shows that for $\alpha<d$
$N_{res,\vec{n}}$ 
diverges with system size $N$. Therefore AT in the Hermitian PLRBM happens at $\alpha_{AT}=d$ independently off $W$ nor $b$.

\textit{Non-Hermitian case}~--~ We generalize the Anderson-Levitov resonance counting to the nH case.
Following the same steps as in the Hermitian one, we subdivide the space in concentric rings around a lattice point $\vec{n}$, Eq.~\eqref{eq:PLRBM_R-ring},
and compare the hopping term $j_R\sim 1/R^{\alpha}$ with the mean level spacing $\delta_R$.
The complex potential, Eq.~\eqref{eq:PLRBM_ep_n_complex}, has a $2$d nature~\cite{DeTomasi2022nonHerm_MBL} and the mean level spacing
is given by the root of the ratio between the entire available area $\sim W^2$ and the number of sites in the ring $\sim R^d$, see Fig.~\ref{fig:nonHerm_resonances}(b),
\be\label{eq:delta_R_nH}
\delta_{R}^{nH} \sim  \frac{W}{R^{d/2}} \ ,
\ee
thus it decays parametrically slower than in the Hermitian case in Eq.~\eqref{eq:delta_R}.
Focusing only on the increase of the level spacing, one might 
conclude that the system is more localized compared to the Hermitian one, and the AT should be at $\alpha_{AT} = d/2$.

\begin{figure}[t!]
\label{fig:Different_W_PLBM}
    \includegraphics[width=1.\columnwidth]{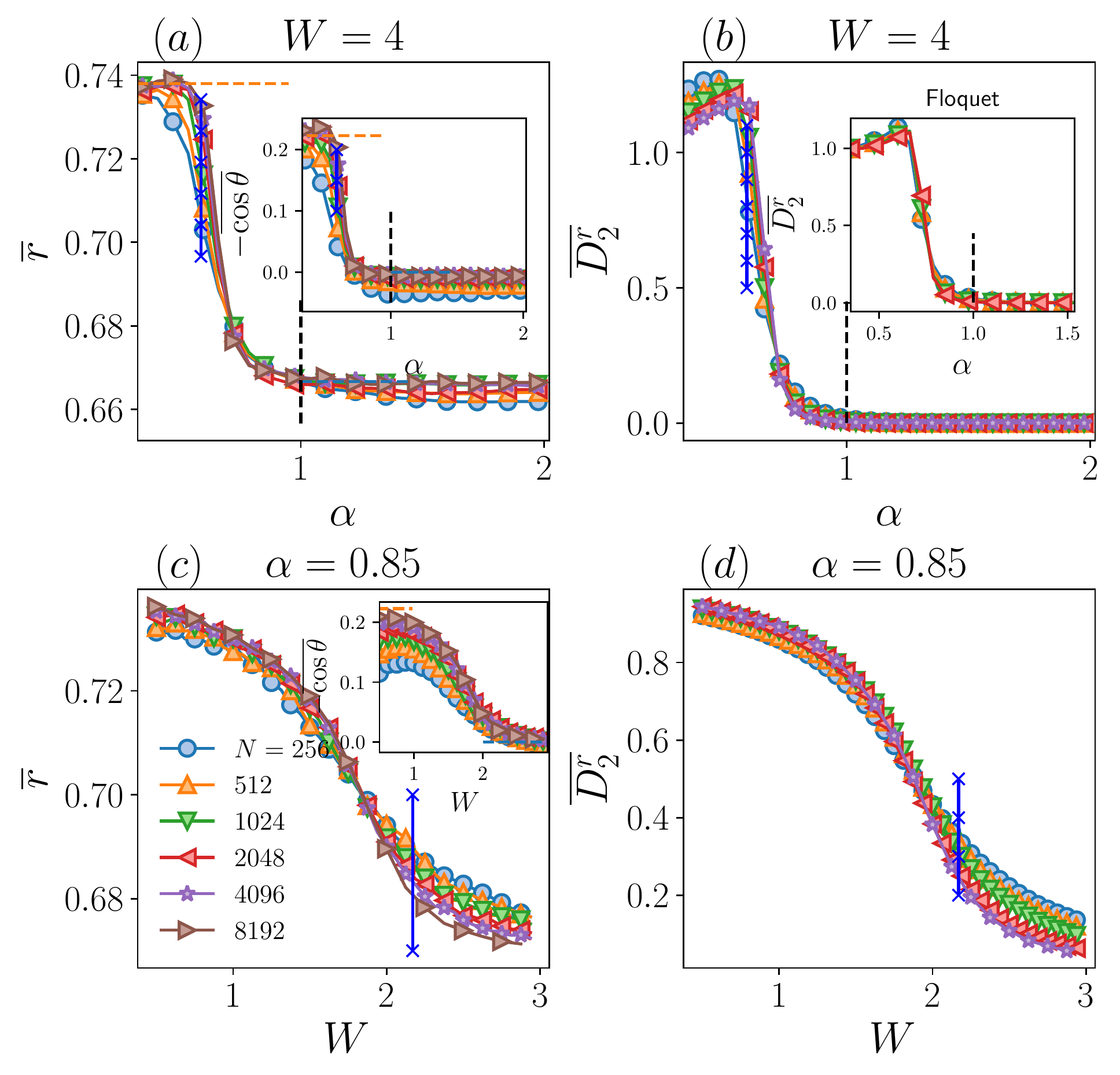}
    \caption{(a),(b)~Spectral statistics and fractal dimension vs $\alpha$ for $W=4$. The vertical dashed lines at $\alpha_{AT}=1$ point AT in the Hermitian case. \rev{(inset to (b)) $\overline{D}_2^r$ for the Floquet model with $W=4$.}
    (c),(d)~$\overline{r}$, $-\overline{\cos{\theta}}$ and $D_q^r$ for fixed $\alpha=0.85$ vs the on-site disorder $W$.  Blue vertical lines are the theoretical predictions,Eq.~\eqref{eq:N_bad_res}, for the critical point with (a,~b)~$N_{bad}(W=4,\alpha_c)=N_c$   (c, ~d)~$N_{bad}(W_c,\alpha =0.85)=N_c$.
    }
\end{figure}

However, as we observe numerically in Figs.~\ref{fig:front_PLBM} and~\ref{fig:Different_W_PLBM}, the transition happens at $W$-dependent value, $d/2\le \alpha_{AT}(W)\le d$.
We show that this is due to the level hybridization, which, in contrast to the Hermitian case, may enhance delocalization by level attraction.
\rev{For clarity, }let's consider the structure of a single resonance for both real and complex $\ep_{\vec{n}}$ and Hermitian hopping $j_{\vec{n}\vec{m}}=j_{\vec{m}\vec{n}}^*$, \rev{because as we show below, the hybridization depends only on the mutual phase of $j_{\vec{m}\vec{n}} j_{\vec{n}\vec{m}}$ and $(\ep_{\vec{n}}-\ep_{\vec{m}})^2$.}

In the resonance condition $|\ep_{\vec{m}}-\ep_{\vec{n}}|<|j_{\vec{n}\vec{m}}|$ for two sites $\vec{n}$ and $\vec{m}$, following the degenerate perturbation theory, one should diagonalize a $2\times2$ matrix
\be
\begin{pmatrix}
\ep_{\vec{m}} & j_{\vec{m}\vec{n}} \\
j_{\vec{n}\vec{m}} & \ep_{\vec{n}} \\
\end{pmatrix},
\ee
leading to the "renormalized" on-site energies
\be\label{eq:Em,En_expression}
\ep_{\vec{m},\vec{n}}^{(1)} = \frac{\ep_{\vec{m}}+\ep_{\vec{n}}}2 \pm \sqrt{\lrp{\frac{\ep_{\vec{m}}-\ep_{\vec{n}}}2}^2+|j_{\vec{m}\vec{n}}|^2}.
\ee

For the Hermitian case, with $\lrp{\ep_{\vec{m}}-\ep_{\vec{m}}}^2\geq 0$, we have level repulsion,
\be\label{eq:Em-En_more_ep_m-ep_n}
\lrv{\frac{\ep_{\vec{m}}^{(1)}-\ep_{\vec{n}}^{(1)}}{\ep_{\vec{m}}-\ep_{\vec{n}}}} =
\sqrt{1+\frac{|j_{\vec{m}\vec{n}}|^2}{\lrp{\ep_{\vec{m}}-\ep_{\vec{n}}}^2}} > 1 \ ,
\ee
leading to the convergent RG for $\alpha>\alpha_{AT}=d$.

In the nH case, the inequality in Eq.~\eqref{eq:Em-En_more_ep_m-ep_n} is not always satisfied. 
For example, at $\lrp{\ep_{\vec{m}}-\ep_{\vec{m}}}^2<0$,
\be\label{eq:Em-En_less_ep_m-ep_n}
\lrv{\frac{\ep_{\vec{m}}^{(1)}-\ep_{\vec{n}}^{(1)}}{\ep_{\vec{m}}-\ep_{\vec{n}}}} = \lrv{\sqrt{1-\lrv{\frac{j_{\vec{m}\vec{n}}}{\ep_{\vec{m}}-\ep_{\vec{n}}}}^2 }} < 1,
\ee
at all $j_{\vec{m}\vec{n}}^2<2\lrv{\ep_{\vec{m}}-\ep_{\vec{n}}}^2$. We refer to this level of attraction as ``bad'' resonances.
It is this new phenomenon of level attraction, Fig.~\ref{fig:nonHerm_resonances}(c), which competes with the enhanced mean level spacing $\delta_{R}^{nH}$ and breaks the RG steps at $d/2<\alpha<\alpha_{AT}(W)$ as soon as the ``bad'' resonance number is significantly large. These ``bad'' resonances appear with a finite (but small) probability per eigenstate, which we next calculate.

We parameterize the hopping $|j_{\vec{m}\vec{n}}|=2 W J$ and diagonal elements with real parameters $J$, $X$, $Y$, $x$, $y$
\be\label{eq:ep_parametrization}
\frac{\ep_{\vec{m}}}W = X + i Y + {x+i y} \ , \quad
\frac{\ep_{\vec{n}}}W = X + i Y - {x+i y} \ ,
\ee
and rewrite the 
condition, Eq.~\eqref{eq:Em-En_less_ep_m-ep_n} in general, as
\be\label{eq:res_cond}
(x^2+y^2) < J^2 < 2(y^2-x^2) \ .
\ee

For a fixed $J$ one can straightforwardly calculate the usual-resonance probability $P_{res}$, defined as the conditional integral $J^2>(x^2+y^2)$ over the parameters $X$, $Y$, $x$, and $y$,
which is present also in the Hermitian case. The ``bad''-resonance one $P_{bad}$, present only in nH case, we calculate via the integral over~\eqref{eq:res_cond}~\footnote{For the exact definitions and expressions of both probabilities see the Supplemental Material~\cite{SM}.}

The number of resonances is given by the summation over the distances $R$ and the integration over the random-amplitude $|s_{\vec{m},\vec{m+R}}|\equiv s$ distribution (being box for $\Re s_{\vec{mn}}$ and $\Im s_{\vec{mn}}$ in Eq.~\eqref{eq:PLRBM_j_mn_box}, with $s\leq 1$):
\be\label{eq:J_vs_s_R^a}
J_{\vec{m},\vec{m+R}} = \frac{|s_{\vec{m},\vec{m+R}}|}{2 W R^\alpha} \ .
\ee
For $W>1/2$, and $d=1$, one obtains the \rev{average }number of the usual and ``bad'' resonances per eigenstate as~\cite{SM}
\begin{multline}\label{eq:N_res_res}
N_{res} = \sum_{\vec{R}}\int P_{res}\lrp{J_{\vec{m},\vec{m+R}}=\frac{s}{2 W R^\alpha}}P(s) ds \\ \revT{\approx} 1.05\frac{ \zeta_{2\alpha}}{W^2} -0.61\frac{\zeta_{3\alpha}}{W^3} + 0.08\frac{\zeta_{4\alpha}}{W^4}
\end{multline}
\begin{multline}\label{eq:N_bad_res}
N_{bad} = \sum_{\vec{R}}\int P_{bad}\lrp{J_{\vec{m},\vec{m+R}}=\frac{s}{2 W R^\alpha}}P(s) ds \\ \revT{\approx}
0.13\frac{\zeta_{2\alpha}}{W^2}-0.09\frac{\zeta_{3\alpha}}{W^3} + 0.01\frac{\zeta_{4\alpha}}{W^4} \ ,
\end{multline}
where $\zeta_x$ is the zeta function, which converges at $x>1$.
As a result, both $N_{res}$ and $N_{bad}$ are finite \rev{and $N$-independent} at $\alpha>d/2$.
If one neglects the  contribution of ``bad'' resonances, Eq.~\eqref{eq:N_res_res} implies that the localization happens at $\alpha>d/2$. However, ``bad'' resonances break down the entire RG analysis by enabling an avalanche of  higher-order resonances.
Therefore, localization in the nH case is achieved only at $N_{bad}\ll 1$ per eigenstate or at $N_{bad}\cdot N \ll N$ resonances in the entire system.
From the numerical results, we find the transition at
\be\label{eq:N_bad_value}
N_{bad}(W,\alpha) \simeq N_c \equiv 0.045 \ ,
\ee
which immediately makes the power-law exponent $\alpha_c(W)$ to be strongly dependent on $W$ via Eq.~\eqref{eq:N_bad_res}. \rev{The average $N_{bad}$ and $N_{res}$ are generically non-integer, meaning that from realization to realization and from eigenstate to eigenstate these integers fluctuate.}

\textit{Numerical results}~--~We restrict our numerics to $d=1$, minimizing finite-size effects. 
To understand the existence of the AT, we study the spectral statistics of the model, Eq.~\eqref{eq:PLRBM_ham}.
In the \revT{extended} phase, we expect the level statistics to be close to that of the Ginibre \revT{ensemble~\cite{Ginibre_ens}}, a nH RMT.
In the localized phase, the energy spectrum should be $2$d Poisson. To separate these two limits, we study the complex gap ratio~\cite{2020_Ribeiro_complex_r-stat}
\begin{equation}
\label{eq:r_statistic_complex}
r_n^{C} = \frac{Z_n^{NN}- Z_n}{Z_n^{NNN} -Z_n},
\end{equation}
where $\{Z_n\}$ is the spectrum of $H$, which is, in general, complex, $Z_n \in  \mathbb{C}$. $Z_n^{NN}$ and $Z_n^{NNN}$ are the nearest neighbor (NN) and the next-nearest neighbor (NNN) of $Z_n$ with respect to the Euclidean distance in $\mathbb{C}$, respectively.
Decomposing $r_n^{C} =r_n e^{i\theta_n}$, we analyze $\{r_n\}$ and $\{\theta_n\}$, separately. For Ginibre RMT $\overline{-\cos{\theta}}\approx 0.229$ and $\overline{r} \approx 0.738$~\cite{2020_Ribeiro_complex_r-stat}, while in the localized phase, we have $\overline{-\cos{\theta}}= 0$ and $\overline{r} = 2/3$. The overline indicates the average over disorder and energy spectrum.
\revT{We have checked that no mobility edge, absent in PLRBM, appears in its non-Hermitian version.}

Figures~\ref{fig:Different_W_PLBM}(a) show $\overline{r}$ and $-\overline{\cos{\theta}}$ as a function of $\alpha$ a fixed $W=4$, respectively.
For small $\alpha$, both quantities approach their RMT prediction (red dashed lines) with increasing $N$, while at large $\alpha$, they tend to the Poisson values.
An abrupt crossover between these limits appears around $\alpha_c\approx 0.72\le \alpha_{AT}=1$, providing evidence of the existence of an AT.
\begin{figure}[t!]
\label{fig:Power_PLBM}
    \includegraphics[width=1.\columnwidth]{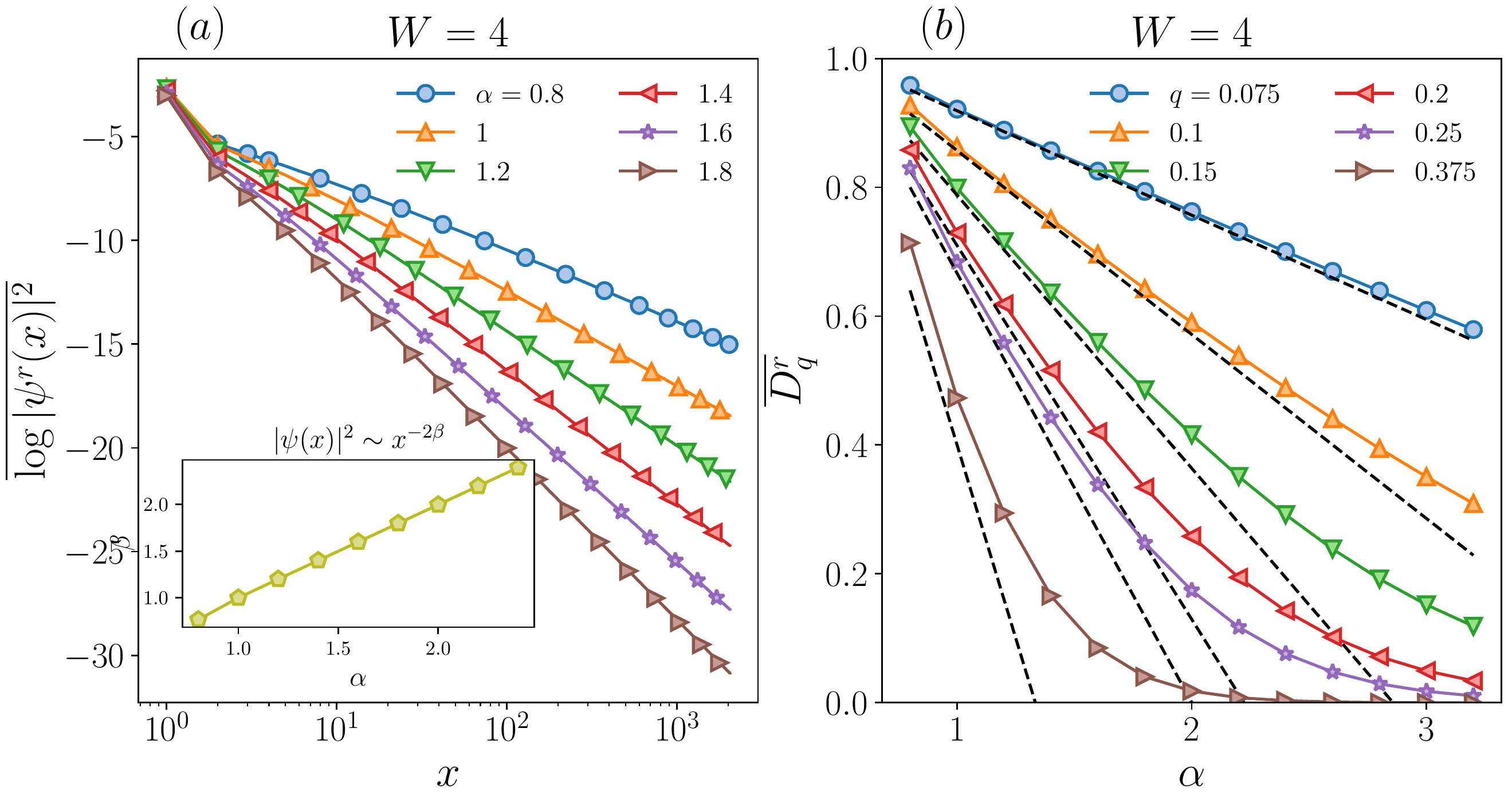}
    \caption{(a)~Eigenstate algebraic spatial decay, $\psi(x) \sim 1/x^{\beta}$, in log-log scale for several $\alpha$ and $N=2048$.
    The inset in (a) shows that $\beta\approx \alpha$.
    (b)~$D_q^r$ vs $\alpha$ for several small $q$. The dashed lines show the analytical prediction $D_q^r= (2\alpha q-1)/(q-1)$.
    }
\end{figure}
To further support the above observation, we compute the inverse participation ratio $IPR_q^r$ and extract a fractal exponent $D_q^r$ via its scaling with the system size $N$
\begin{equation}
    IPR_q^r = \sum_{\vec{m}}|\psi_n^r(\vec{m})|^{2q}, \;\; D_q^r = \frac{1}{1-q}\frac{d\log{IPR_q^r}}{d\log{N}}.
\end{equation}
Here $\psi_n^r$ is the right eigenvector with eigenvalue $Z_n$~\footnote{We have checked in the Supplemental Materials~\cite{SM} that $IPR$ is statistically the same for left eigenvectors and mixed definitions  $IPR_q^r = \sum_{\vec{m}}|\psi_n^r(\vec{m})\psi_n^l(\vec{m})|^{q}$.}.
$0\leq D_q^r\leq 1$ measures the spread of the wave functions. In the ergodic phase, $D_q^r\to 1$, while in the power-law localized phase, the behavior of $D_q^r$ depends on the power $\beta$ of the decay of the wave function $\psi(\vec{m})\sim 1/|\vec{m}|^{\beta}$: $D_q^r=0$ for $q>1/(2\beta)$, while for $q<1/(2\beta)$ the fractal exponent is $D_q^r=(2\alpha q-1)/(q-1)$.
Varying $q$ in $D_q^r$, we can probe if the phase is power-law localized.
As shown in Fig.~\ref{fig:Different_W_PLBM}~(b), $D_2^r\approx 1$ for small $\alpha<0.85$ indicates ergodicity. For larger $\alpha$, we observe that $D_2^r\to 0$. Importantly, $D_2^r\to 0$ also at $\alpha_c<\alpha<d=1$.

Now, 
to understand the dependence of the critical point on the onsite-disorder strength, we tune $W$.
Figure~\ref{fig:Different_W_PLBM}(c)-(d) shows $\overline{r}$, $-\overline{\cos{\theta}}$ and $\overline{D_2^r}$ at $\alpha=0.85<\alpha_{AT}$ versus $W$. The system-size $N$ increase tends the curves in Fig.~\ref{fig:Different_W_PLBM}~(c)-(d) to their ergodic values at $W\le 1.85$, while we observe localization at larger $W$. The dependence of the critical point $d/2\le \alpha_c\le d$ on the disorder strength $W$ is in a good agreement with the theoretical prediction of Eqs.~\eqref{eq:N_bad_res},~\eqref{eq:N_bad_value}, see Fig.~\ref{fig:front_PLBM} and Fig.~\ref{fig:Different_W_PLBM}. In particular, we have that $\alpha_c\rightarrow 1/2$ as $1/W^2$ at large $W$, Fig.~\ref{fig:front_PLBM}.

\rev{We tested our finding also in the Floquet case, see the inset of Fig.~\ref{fig:Different_W_PLBM}(b). As a result, the non-unitary Floquet model undergoes a localization-delocalization transition at $\alpha_c(W)\le \alpha_{AT}$. 
Using the relation between entanglement entropy and transport properties for Gaussian state, we predict the existence of measurement-induced transition in entanglement in that model.}

Having provided the evidence 
of the $W$-dependence of $\alpha_c$, we turn to the type of localization. Within the localized phase, we analyze the eigenstate spatial decay from its absolute-value maximum~\cite{Deng2018duality,Nosov2019correlation,Nosov2019mixtures,deng2020anisotropy,Kutlin2020_PLE-RG,Motamarri2021RDM,Tang2022nonergodic}.
The wave-function decay is algebraic, i.e. $|\psi(x)|^2\sim 1/|x|^{2\beta}$, Fig.~\ref{fig:Power_PLBM}(a), and the power $\beta$ is given by $\beta\approx \alpha$ in agreement with the nH perturbation theory at $\alpha_c<\alpha<d=1$, see inset of Fig.~\ref{fig:Power_PLBM}(a).
The algebraic wave-function profile also manifests itself in $D_q^r$ vs $\alpha$ for several $q$, see Fig.~\ref{fig:Power_PLBM}(b). In good agreement with the theoretical prediction for power-law localized phases, $D_q^r\approx (2\alpha q-1)/(q-1)$ for $q<1/(2\alpha)\approx 1/(2\beta)$ and zero otherwise.

\textit{Conclusion}~--~In this work, we inspect the fate of the Anderson transition in a nH PLRBM model with random gain and loss in $d$ dimensions. We provide numerical and analytical evidence of the existence of AT, which happens at a smaller power $\alpha$ than in the Hermitian counterpart.

We generalize the Anderson-Levitov resonance counting to the nH case to elaborate an analytical understanding of the system. This technique reveals the emergence of two competing mechanisms: the parametric enhancement of the level spacing, enhancing localization, and the emerging ``bad'' resonances, favoring delocalization.
The competition between the two leads to the Anderson transition at $d/2\leq \alpha_c(W) \leq d$.
The critical point depends on the on-site-disorder amplitude $W$, similar to short-range models in $d>2$. This result should be compared to the Hermitian case, where the transition happens at $\alpha=d$ for any $W$.
We also show that in the localized phase, the wave-function envelopes decay algebraically from their maxima. The spatial decay coincides with the one of the PLRBM hopping term, also for $\alpha_c(W) \leq \alpha\leq d$.
As a result, nH systems might also host localized algebraic phases with a decay rate lower than $d$, which is forbidden in Hermitian systems.

\revT{Together with the UM, nH in PLRBM opens the direct applications to the avalanche theory of many-body delocalization, suggested in~\cite{Suntajs2023similarity} for the Hermitian case, and should stabilize the MBL phase beyond its Hermitian limit.~\cite{DeTomasi2022nonHerm_MBL}}


\textit{Acknowledgments—}~--~We thank \rev{D.~Belkin, B.~Clark}, T.~J.~Hughes and V.~E.~Kravtsov for illuminating discussions.
I.~M.~K. acknowledges the support
from the European Research Council under the European Union's Seventh Framework Program Synergy ERC-2018-SyG HERO-810451.
G.~D.~T. acknowledges the support from the EPiQS Program of the Gordon and Betty Moore Foundation.

\bibliography{Lib}

\begin{widetext}
\appendix

\section{Calculation of the probabilities of resonances}\label{App:Res_integrals}
Using the parametrization~(13) 
and the left condition in~(14), 
for the box-distributed $\Re \ep_{\vec{n}}$ and $\Im \ep_{\vec{n}}$ in the interval $\lrv{\Re \ep_{\vec{n}}}, \lrv{\Im \ep_{\vec{n}}}< W$, Eq.~(2b), 
we have
\be
|X|<1-{|x|} \ , \quad
|Y|<1-{|y|} \
\ee
and the $j$-resolved probability of usual resonances
\begin{multline}\label{SM:P_res}
P_{res}(j) = \iiiint\limits_{-W}^{W} \theta\lrp{|j|-|\ep_{\vec{m}}-\ep_{\vec{n}}|}\frac{d\Re \ep_{\vec{n}}}{2W} \frac{d\Im \ep_{\vec{n}}}{2W} \frac{d\Re \ep_{\vec{m}}}{2W} \frac{d\Im \ep_{\vec{m}}}{2W} =
\iint\limits_{-1}^{1} \lrp{1-|x|}\lrp{1-|y|}\theta\lrp{J^2-x^2-y^2}dx dy =\\=
\left\{
\begin{array}{ll}
\int\limits_{0}^1 (1-x)\lrb{2\sqrt{J^2 - x^2}-\lrp{J^2 - x^2}} dx, & J<1 \\
\int\limits_{\sqrt{J^2-1}}^1 (1-x)\lrb{2\sqrt{J^2 - x^2}-\lrp{J^2 - x^2}} dx +
\int\limits_0^{\sqrt{J^2-1}}(1-x)dx, & 1<J<\sqrt{2} \\
1, & J>\sqrt{2}\\
\end{array}
\right.
=\\=
\left\{
\begin{array}{ll}
\pi J^2 -\frac{8 J^3}{3} + \frac{J^4}{2}, & J<1 \\
\frac13 + \frac43(1+2J^2)\sqrt{J^2-1} + J^2\lrb{\pi-2-4\arccos\lrp{\frac1J}}-\frac{J^4}{2}, & 1<J<\sqrt{2} \\
1, & J>\sqrt{2}\\
\end{array}
\right.
\end{multline}
with $J = |j|/(2W)$.

\revT{Note here that the choice of the real and imaginary amplitudes to be the same, $\lrv{\Re \ep_{\vec{n}}}, \lrv{\Im \ep_{\vec{n}}}< W$, is arbitrary and does not change much.
The difference between the amplitudes of real $\lrv{\Re \ep_{\vec{n}}}<W_R$ and imaginary $\lrv{\Im \ep_{\vec{n}}}<W_I$ parts of disorder makes only quantitative difference in the phase diagram, which is straightforwardly seen from the above Eq.~\eqref{SM:P_res}.
Indeed, as soon as both real and imaginary parts do not scale with the system size, the qualitative results are the same: the localization transition is shifted to smaller values of the power $\alpha$ and the critical line $\alpha_c(W_R, W_I)$, unlike the Hermitian case, depends on the disorder amplitudes $W_R$, $W_I$.
}

\begin{figure}[t!]
    \includegraphics[width=0.3\columnwidth]{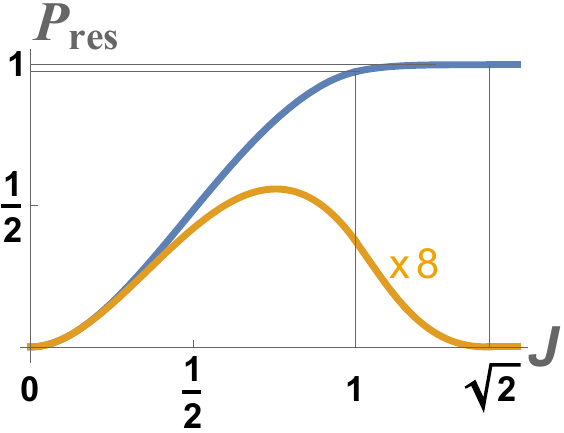}
    \caption{Probabilities of (blue)~usual $P_{res}$ and (orange)~``bad'' $P_{bad}$ resonances as functions of $J=|j_{\vec{m}\vec{n}}|/(2W)$. The latter is multiplied by $8$.}
\label{Fig:Pres}
\end{figure}


The plot of the above formula~\eqref{SM:P_res} is given in Fig.~\ref{Fig:Pres} (blue line) of the main text.
One can see that the probability of usual resonances monotonically increases towards $1$ with increasing hopping term $J=|j_{\vec{m}\vec{n}}|/(2W)$ with respect to the disorder amplitude $W$.

The corresponding probability $P_{res}^0(j)$ for the Hermitian case of $Y\equiv y\equiv 0$ as
\be\label{SM:P_res_Herm}
P_{res}^0(j) = \iint\limits_{-W}^{W} \theta\lrp{|j|-|\ep_{\vec{m}}-\ep_{\vec{n}}|}\frac{d\ep_{\vec{n}}}{2W} \frac{d\ep_{\vec{m}}}{2W} =
\int\limits_{-1}^{1} \lrp{1-|x|}\theta\lrp{J^2-x^2}dx=
\left\{
\begin{array}{ll}
2 J - J^2, & J<1 \\
1, & J>1\\
\end{array}
\right.
\ee

The probability of ``bad'' resonances is given by the conditions~(12) and~(14)
and takes the form

\begin{multline}\label{SM:P_bad}
P_{bad}(j) = \iiiint\limits_{-W}^{W} \theta\lrp{|j|-|\ep_{\vec{m}}-\ep_{\vec{n}}|}\theta\lrp{1-\lrv{1+\frac{|j|^2}{\lrp{\ep_{\vec{m}}-\ep_{\vec{n}}}^2}}}
\frac{d\Re \ep_{\vec{n}}}{2W} \frac{d\Im \ep_{\vec{n}}}{2W} \frac{d\Re \ep_{\vec{m}}}{2W} \frac{d\Im \ep_{\vec{m}}}{2W} =\\=
\iint\limits_{-1}^{1} \lrp{1-|x|}\lrp{1-|y|}\theta\lrp{J^2-x^2-y^2}\theta\lrp{y^2-x^2-\frac{J^2}2}dx dy
=\\=
\left\{
\begin{array}{ll}
\int\limits_{0}^1 (1-x)\lrb{2\sqrt{J^2 - x^2}-\lrp{J^2 - x^2} - 2\sqrt{\frac{J^2}2 + x^2}+\frac{J^2}2 + x^2} dx, & J<1 \\
\int\limits_{\sqrt{J^2-1}}^1 (1-x)\lrb{2\sqrt{J^2 - x^2}-\lrp{J^2 - x^2}- 2\sqrt{\frac{J^2}2 + x^2}+\frac{J^2}2 + x^2} dx +\atop+
\int\limits_0^{\sqrt{J^2-1}}(1-x)\lrp{1- 2\sqrt{\frac{J^2}2 + x^2}+\frac{J^2}2 + x^2}dx, & 1<J<\frac{2}{\sqrt{3}} \\
\int\limits_0^{\sqrt{1-\frac{J^2}2}}(1-x)\lrp{1- 2\sqrt{\frac{J^2}2 + x^2}+\frac{J^2}2 + x^2}dx, & \frac{2}{\sqrt{3}}<J<\sqrt{2} \\
0, & J>\sqrt{2}\\
\end{array}
\right.
=\\=
\left\{
\begin{array}{ll}
\lrp{\frac{\pi}3-\frac12\ln\lrp{2+\sqrt{3}}}J^2+
\lrp{\sqrt{3}-\frac{5+\sqrt{2}}3}J^3 + \frac{J^4}{16}, & J<1 \\
\frac16 + \frac23 (1 + 2 J^2) \sqrt{J^2 - 1} + J^2 \lrb{\frac{\pi}3 - 1 - \ln\lrp{\frac{1 + \sqrt{3}}{\sqrt{2}}}- 2\arccos\lrp{\frac1J}} +
\lrp{\sqrt3 - \frac{1 + \sqrt2}3}J^3 - \frac{7 J^4}{16}, & 1<J<\frac{2}{\sqrt{3}} \\
-\frac16 + \frac{1 + J^2}3 \sqrt{4 - 2 J^2} + \frac{J^2}2 -
 \frac{\sqrt{2}}3 J^3 + \frac{J^4}8 - J^2 \ln\lrp{\frac{\sqrt{2}+\sqrt{2 - J^2}}J}, & \frac{2}{\sqrt{3}}<J<\sqrt{2} \\
0, & J>\sqrt{2}\\
\end{array}
\right.
\ .
\end{multline}
The above expression is plotted in Fig.~\ref{Fig:Pres} (orange line) in the main text. 
Unlike the case of the usual resonances, the probability of ``bad'' ones is non-monotonic and has the maximum at $J\simeq 0.7554$.

These $J$-resolved probabilities can be used for the calculation of the number of resonances by using the expressions Eq. (13)
and the parametrization $J_{\vec{m},\vec{m+R}} = {|s_{\vec{m},\vec{m+R}}|}/\lrp{2 W R^{\alpha}}$,~Eq.~(15). 
\bes\label{SM:N_res,bad}
\begin{align}
N_{res} &= \sum_{\vec{R}}\int P_{res}\lrp{J_{\vec{m},\vec{m+R}}=\frac{s}{2W R^\alpha}}P(s) ds \\
N_{bad} &= \sum_{\vec{R}}\int P_{bad}\lrp{J_{\vec{m},\vec{m+R}}=\frac{s}{2W R^\alpha}}P(s) ds \ ,
\end{align}
\ees
where the summation over $\vec{R}$ is given over the $\mathbb{Z}^d$-lattice and the averaging over $s$ is taken over the random amplitudes of $s=|j_{\vec{m},\vec{m+R}}|R^{\alpha}$.

From the comparison of the expressions~\eqref{SM:P_res} and~\eqref{SM:P_res_Herm} one can immediately see that the summation over $R$ will diverge at large $R$ (small $J$) due to the first terms in the above expressions.
It is this difference in powers of small-$J$ expansion $P_{res}^0 \sim J$ and $P_{res}\sim J^2$ which shifts the divergence from $\alpha=d$ in the Hermitian case to $\alpha\to d/2$ (in the absence of ``bad'' resonances) in the nH case.

Both numbers of usual and ``bad'' resonances in the range $J>1$ (i.e. at $W>1/2$ for all $R\geq 1$ and $s<1$)~\eqref{SM:N_res,bad} take the following form
\bes\label{eq:SM_N_res_bad}
\begin{align}
N_{res} &= \frac{\pi \zeta_{2\alpha}}{W^2}\mean{s^2} -\frac{8\zeta_{3\alpha}}{3W^4}\mean{s^3} + \frac{\zeta_{4\alpha}}{2W^4}\mean{s^4}\\
N_{bad} &=
\lrp{\frac{\pi}3-\frac12\ln\lrp{2+\sqrt{3}}}\frac{\zeta_{2\alpha}}{W^2}\mean{s^2}+\lrp{\sqrt{3}-\frac{5+\sqrt{2}}3}\frac{\zeta_{3\alpha}}{W^3}\mean{s^3} + \frac{\zeta_{4\alpha}}{16 W^4}\mean{s^4} \ ,
\end{align}
\ees
where $\zeta_x$ is a zeta function, which converges at $x>1$.

In order to find the moments $\mean{s^q}$, $2\leq q\leq 4$, one should calculate its distribution.
For the box distribution of $|\Re j|$ and $|\Im j|$, Eq.~(2b), 
with the real $s_R$ and imaginary $s_I$ parts of $s = s_R + i s_I$ in the interval
$|s_R|, |s_I|<1/\sqrt{2}$,
one can find the one of $s$ as
\be
P(s) = \iint_{-1}^1 \frac{ds_R ds_I}{4}\delta(s-\sqrt{s_R^2+s_I^2}) =
s\lrb{\pi - 4\arccos\lrp{\frac{1}{\sqrt{2} s}}\theta\lrp{s-\frac{1}{\sqrt{2}}}}\theta\lrp{1-s} \ .
\ee
The corresponding moments are given by
\be
\mean{s^2} = \frac13 \ , \quad
\mean{s^3} = \frac{7}{40}+\frac{\ln 2+6 \ln (1+\sqrt{2})}{80\sqrt{2}} \simeq 0.23 \ , \quad
\mean{s^4} = \frac{7}{45}
\ee

After the substitution of $\mean{s^q}$ to
the above expressions~\eqref{eq:SM_N_res_bad}, one obtains
\bes\label{eq:SM_N_res_bad_result}
\begin{align}
N_{res} &= \frac{\pi \zeta_{2\alpha}}{3W^2} -\frac{\zeta_{3\alpha}\lrb{14\sqrt{2}+\ln 2+6\ln\lrp{1+\sqrt{2}}}}{30\sqrt{2}W^4} + \frac{7\zeta_{4\alpha}}{90W^4}\\
N_{bad} &=
\lrp{\frac{\pi}3-\frac12\ln\lrp{2+\sqrt{3}}}\frac{\zeta_{2\alpha}}{3W^2}+\lrp{\sqrt{3}-\frac{5+\sqrt{2}}{3}}\lrb{\frac{\lrb{14\sqrt{2}+\ln 2+6\ln\lrp{1+\sqrt{2}}}}{80\sqrt{2}}}\frac{\zeta_{3\alpha}}{W^3} + \frac{7\zeta_{4\alpha}}{720 W^4} \ ,
\end{align}
\ees
coincide with the approximate expressions~(16) and (17) in the main text. 
The expressions for $W<1/2$ are straightforward to calculate from~\eqref{SM:N_res,bad},~\eqref{SM:P_res}, and~\eqref{SM:P_bad}, but have quite cumbersome analytical expressions.

Like in the Hermitian case at $\alpha<d$, where the presence of the extensive number of resonances $N_{res}$ per eigenstate breaks down the spatial renormalization group~\cite{Levitov1990}, in the nH case the presence of a small fraction of ``bad'' resonances per eigenstate does this job.

Indeed, the spatial renormalization group~\cite{Levitov1990} is based on the assumption that one can hybridize the resonance pairs one-by-one, going from the strongest ones (at smaller distances) to the weaker ones (at larger distances) and never coming back to the same (previously resonant) pair, due to its hybridization.

In the nH ``bad'' resonances the situation is drastically different as each of such resonances attracts the levels even closer via the hybridization. This opens the possibility of the recurring resonance pairs at later stages of the renormalization group and, thus, to the avalanche of resonances.
In order to avoid having such avalanche effects, one needs to have the number of \rev{``bad''} resonances per eigenstate small enough.
We estimate this number numerically in the main text.
\subsection{Numerical results}
In this section, we show further numerical data to support our claims. In particular, in the main text, we consider the inverse participation (IPR) ratio for the right eigenvector. Now, we consider the more generally defined as
\begin{equation}
\label{eq:IPR_left_right}
IPR_2 = \sum_{\vec{m}} |\psi^r_n(\vec{m})|^2|\psi^l_n(\vec{m})|^2,
\end{equation}
where $\psi^r_n$ and $\psi^l_n$ are the right- and left- eigenvectors. $D_2$ is the respective fractal exponent
\begin{equation}
\label{eq:D_left_right}
D_2 = -\frac{d\log{IPR_2}}{d\log{N}}.
\end{equation}
Figures~\ref{fig:PLBM_left_right}(a)-(b) show $IPR_2$ and $D_2$. These panels should be compared to Figs.~\ref{fig:Specturm_PLBM}(c)-(d),
where we computed for the same value of $W$ and $b$ the right $IPR^r_2$ and $D_2^r$.  We do not observe any substantial changes and $D_2^r\approx D_2$.
\begin{figure}[t!]
\label{fig:PLBM_left_right}
    \includegraphics[width=0.6\columnwidth]{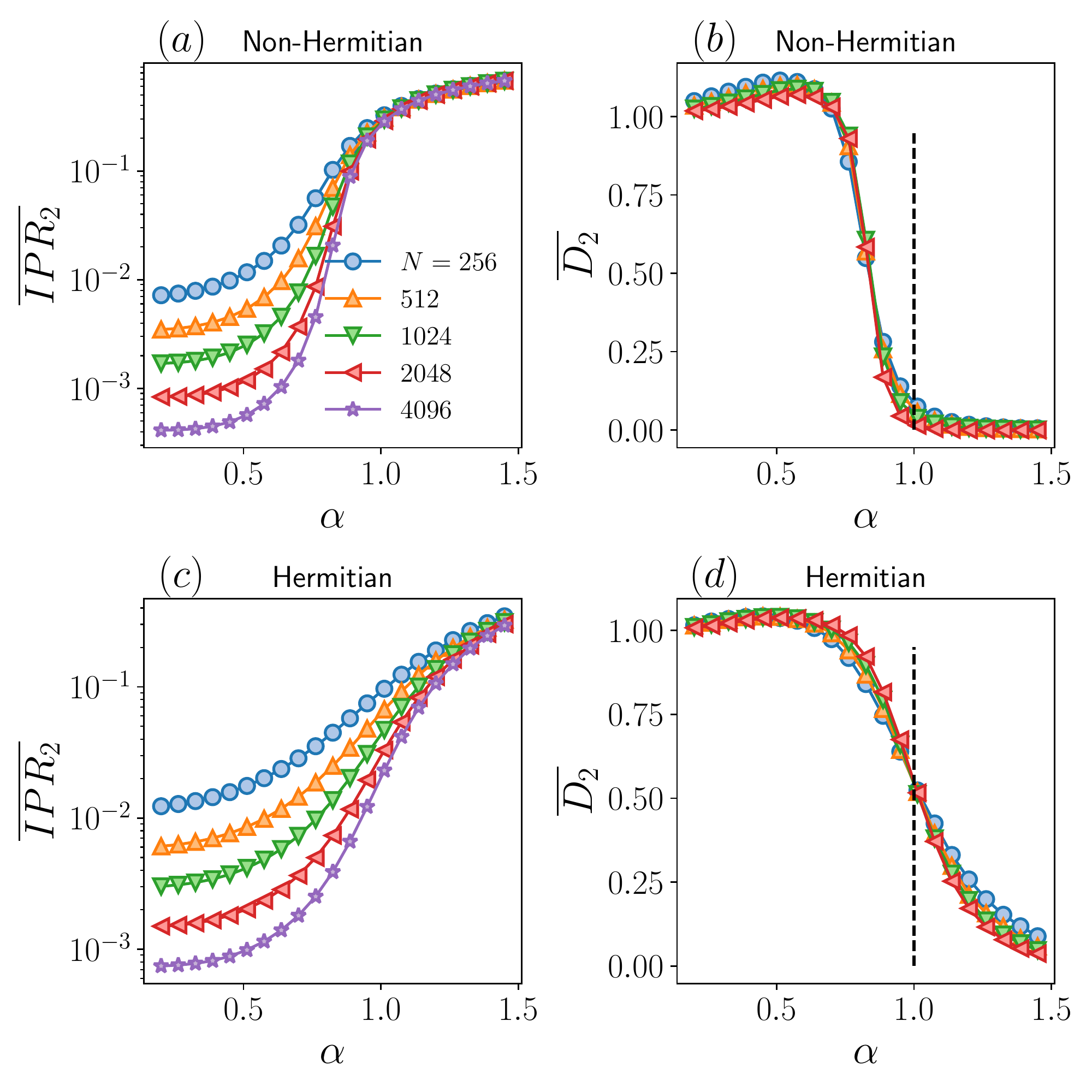}
    \caption{(a),(b)~$\overline{IPR}_2$ and $\overline{D}_2$ vs $\alpha$ for $W=2$ and several $N$ for the nH model.
    (c),(d)~$\overline{IPR_2}$ and the respective fractal dimension $\overline{D_2}$ for $W=4$ for the Hermitian case.
    The vertical dashed line $\alpha_{AT}=1$ points the phase transition in the Hermitian case.
    }
\end{figure}

\begin{figure}[t!]
\label{fig:Specturm_PLBM}
    \includegraphics[width=0.6\columnwidth]{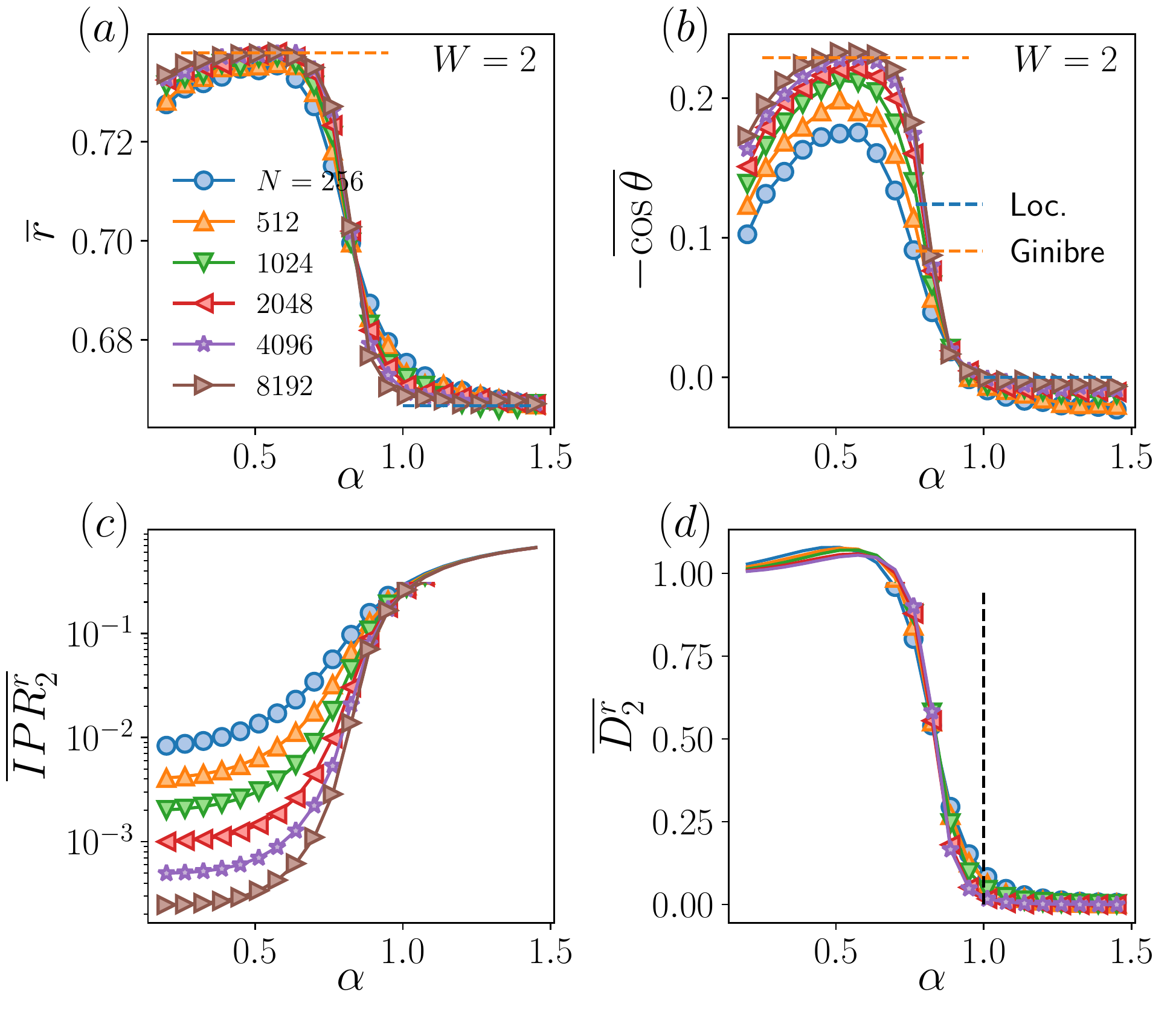}
    \caption{\rev{Level spacing ratio $r_n^{C} \equiv \lrp{Z_n^{NN}- Z_n}/\lrp{Z_n^{NNN} -Z_n}=r_n e^{i\theta_n}$ statistics: (a),(b)~Averaged magnitude $\overline{r}$ and the cosine of the phase $-\overline{\cos{\theta}}$ of spectral statistics} vs $\alpha$ for $W=2$ and several $N$.
    (c),(d)~$\overline{IPR_2}$ and the respective fractal dimension $\overline{D_2^r}$ for $W=2$.
    The vertical dashed line $\alpha_{AT}=1$ points the phase transition in the Hermitian case.
    }
\end{figure}

\begin{figure}[t!]
\label{fig:Floquet_PLBM}
    \includegraphics[width=0.6\columnwidth]{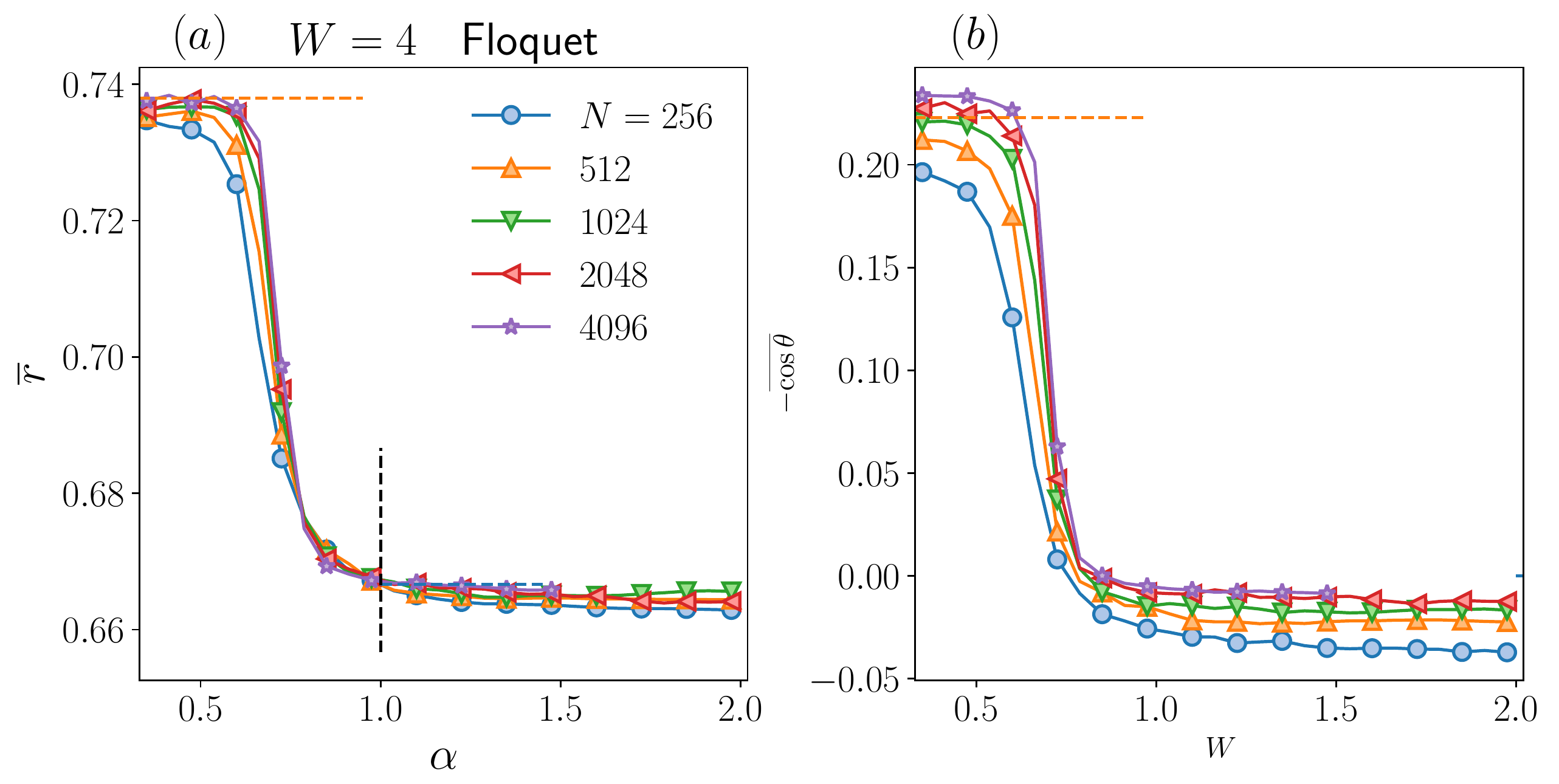}
    \caption{\rev{Level spacing ratio $r_n^{C} \equiv \lrp{Z_n^{NN}- Z_n}/\lrp{Z_n^{NNN} -Z_n}=r_n e^{i\theta_n}$ statistics: (a),(b)~Averaged magnitude $\overline{r}$ and the cosine of the phase $-\overline{\cos{\theta}}$ of spectral statistics vs $\alpha$ for $W=4$ and several $N$, for the Floquet model.}
    }
\end{figure}

Furthermore, in Fig.~\ref{fig:PLBM_left_right}, we show $IPR_2$ and $D_2$ for the Hermitian case. For the Hermitian case, we keep the diagonal disorder strength having the same magnitude as the absolute magnitude of the nH one. The Hermitian case is more delocalized compared to its nH counterpart. At the Anderson critical point $\alpha_{AT}=1$, the fractal dimension shows multifractality ($0<D_2<1$) for the Hermitian case, while for the nH one $D_2\approx 0$,
see dashed black for reference.

Figure~\ref{fig:Specturm_PLBM} shows further results for nH cases with $W=2$. The Panels (a)-(b) focus on the energy spectrum, while the (c)-(d) on the $IPR^r_2$ and its fractal dimension $D_2^r$. In agreement with the results in the main text,  we observe a cross-over from ergodic to localized for $\alpha_c\approx 0.85<\alpha_{AT}=1$.

\rev{Finally, in Fig.~\ref{fig:Floquet_PLBM}~(a)-(b), we show the spectral statistic for the non-Hermitian Floquet Hamiltonian
$H^{F} = -\log{U}$, with $U = e^{-i\revT{T}\sum_{\vec{mn}} j_{\vec{nm}} c^\dagger_\vec{m} c_\vec{n}} e^{-\revT{T}\sum_{\vec{n}} \ep_{\vec{n}} c^\dagger_\vec{n} c_\vec{n}}$\revT{, with a period $T$, set to be $1$}. Like the fractal dimension, shown in the main text, the Floquet model present a transition for $\alpha_c(W)<\alpha_{AT}$}.

\end{widetext}

\end{document}